\documentstyle[12pt]{article}
\begin{document}

\title{QCD Anomaly Coupling for the $\eta'-g-g$ Vertex in Inclusive
       Decay $B\to\eta' X_s$}
\author{ Taizo Muta and Mao-Zhi Yang\\
       Department of Physics, Hiroshima University, Higashi-Hiroshima,\\
       Hiroshima 739-8526, Japan{\thanks{mailing address}}}
\maketitle
\begin{center}
\begin{minipage}{120mm}
\vskip 0.8in
\begin{center}{\bf Abstract} \end{center}
{The QCD anomaly coupling of $\eta'-g-g$ is treated as the axial vector
current triangle anomaly. By assuming the divergent axial vector coupling
of the $\eta'$ meson with the quark line in 
the triangle diagram, we calculate the QCD anomaly of 
$\eta'-g-g$ as well as the QED 
anomaly of $\eta'-\gamma-\gamma$ with only one common parameter 
$\kappa_{\eta'}$. We obtain consistent results of the branching
ratios for $B\to\eta' X_s$, $J/\psi\to\eta'\gamma$ and 
$\eta'\to \gamma\gamma$ when comparing with experimental data.}
\end{minipage}
\end{center}
\vskip 1in

\newpage

The first observation of $B\to\eta' X_s$ decays \cite{talk,1} has
stimulated much theoretical 
interests. A number of interpretations have been proposed to 
explain the observed large decay branching fraction. One of the 
interpretations
is based on $b\to c\bar{c}s$ decay with the assumption 
of the intrinsic $c\bar{c}$ 
content in the $\eta'$ wave function\cite{2}, while the others are
based on the $b\to sg^*$
penguin transition followed by $g^*\to\eta' g$ through QCD anomaly
\cite{3,4,5,6,7,8}. The observed recoil mass spectrum favors
the second interpretation\cite{1}. Thus the QCD anomaly coupling of
the $\eta'$ meson with gluon fields is important in the $B\to\eta' X_s$
decay process. However there are still no consensus among the authors
on the second interpretation. Some of them conclude that the calculation
within the standard model (SM) is sufficient to account for the 
experimental data\cite{3,6,8} while the others\cite{4,5} conclude not,
stressing that new physics is needed for the proper interpretation. 
The key point of the confusing situation
comes from the QCD anomaly itself. Refs.\cite{3,4,6} stress that 
the QCD 
anomaly coupling for the $\eta'-g-g$ vertex is highly nonperturbative
so that it 
is unpredictable. Because the $\eta'-g-g$ vertex 
is treated differently in
Refs.\cite{3,4,5,6}, the different results of the branching ratio of
$B\to \eta' X_s$ are drawn.

In the present paper, we try to make clear the above 
confusing situation on the $\eta'-g-g$ 
coupling. Given the unique properties of the axial current
triangle anomaly which has first been discovered in QED and successfully
solved the problem of the $\pi^0\to\gamma\gamma$ decay\cite{9} 
the $\eta'-g-g$ 
coupling is also treated as the axial current anomaly here (Fig.1). 
The vertex
of the $\eta'$ meson coupling with the quark line in the triangle
diagram is assumed to be given by 
$i\kappa_{\eta'}/ {\hskip -2.4mm p\gamma_5}$
which comes from the PCAC hypothesis\cite{10}. Here $\kappa_{\eta'}$
is introduced as a coupling parameter. It is resonable to treat 
$\kappa_{\eta'}$ as a constant, because it only involves the inner 
properties of the $\eta'$ meson.

\begin{figure}
\setlength{\unitlength}{0.13in} 
\begin{center}
\begin{picture}(35,10)
\put(15,5){\line(-1,0){2}}
\put(13,5){\circle*{1}}
\put(13,5){\line(-2,1){6}}
\put(10,6.5){\vector(-2,1){0.1}}
\put(7,8){\line(0,-1){6}}
\put(7,5){\vector(0,-1){0.1}}
\put(7,2){\line(2,1){6}}
\put(10,3.5){\vector(2,1){0.1}}

\multiput(6.3,7.8)(-1,0){6}{\oval(1.4,1.4)[t]}
\multiput(5.8,7.8)(-1,0){5}{\oval(0.4,0.5)[b]}
\multiput(6.3,2.)(-1,0){6}{\oval(1.4,1.4)[t]}
\multiput(5.8,2.)(-1,0){5}{\oval(0.4,0.5)[b]}

\put(0,8.8){$k_1$}
\put(2.5,9){\vector(1,0){1}}
\put(0,3.3){$k_2$}
\put(1.5,3.3){\vector(1,0){1}}
\put(3.5,4.7){$q{\hskip -0.5mm+}{\hskip -0.mmk_1}$}
\put(10,7.5){$q$}
\put(10,2.5){$q{\hskip -0.5mm+}{\hskip -0.5mmp}$}
\put(14,4){$p$}
\put(15,4.5){\vector(1,0){1}}

\put(5,9){$igT_a\gamma^{\mu}$}
\put(5,0.5){$igT_b\gamma^{\nu}$}
\put(12.5,6){$i/{\hskip -2.4mm p}\gamma_5\kappa_{\eta'}$}
\put(33,5){\line(-1,0){2}}
\put(31,5){\circle*{1}}
\put(31,5){\line(-2,1){6}}
\put(28,6.5){\vector(-2,1){0.1}}
\put(25,8){\line(0,-1){6}}
\put(25,5){\vector(0,-1){0.1}}
\put(25,2){\line(2,1){6}}
\put(28,3.5){\vector(2,1){0.1}}

\multiput(24.3,7.8)(-1,0){6}{\oval(1.4,1.4)[t]}
\multiput(23.8,7.8)(-1,0){5}{\oval(0.4,0.5)[b]}
\multiput(24.3,2.)(-1,0){6}{\oval(1.4,1.4)[t]}
\multiput(23.8,2.)(-1,0){5}{\oval(0.4,0.5)[b]}

\put(18,8.8){$k_2$}
\put(19.5,9){\vector(1,0){1}}
\put(18,3.3){$k_1$}
\put(19.5,3.3){\vector(1,0){1}}
\put(21.5,4.7){$q{\hskip -0.5mm+}{\hskip -0.mmk_2}$}
\put(28,7.5){$q$}
\put(28,2.5){$q{\hskip -0.5mm+}{\hskip -0.5mmp}$}
\put(32,4){$p$}
\put(33,4.5){\vector(1,0){1}}

\put(23,9){$igT_b\gamma^{\nu}$}
\put(23,0.5){$igT_a\gamma^{\mu}$}
\put(30.5,6){$i/{\hskip -2.4mm p}\gamma_5\kappa_{\eta'}$}
\end{picture}
\end{center}
\caption{The lowest-order diagrams which contribute 
       to QCD anomaly coupling of $\eta'-g-g$. Higher order diagrams
       do not contribute to the anlmaly term according to the 
       Adler-Bardeen theorem\cite{adler} }
\end{figure}
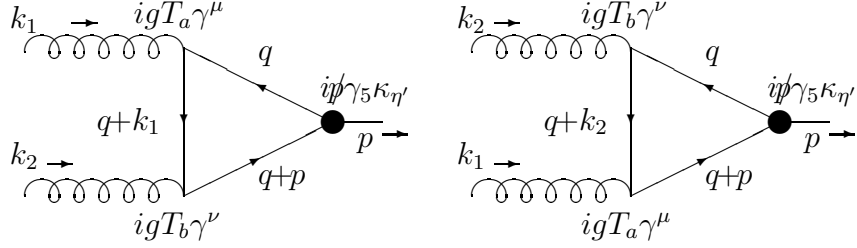

The routing of the loop momentum $q$ is shown in Fig.1. We write the 
amplitude $T^{\mu\nu\lambda}$ directly by following the Feynman rule,
\begin{eqnarray}
 T^{\mu\nu\lambda}=\int \frac{d^4q}{(2\pi)^4}(-1)\left\{Tr\left[
          \frac{i}{/{\hskip -2.4mm q}-m}i\gamma^{\lambda}\gamma_5 
          \kappa_{\eta'}\frac{i}{/{\hskip -2.4mm q}+/{\hskip -2.4mm p}-m}
          ig\gamma^{\nu}T_b\frac{i}{/{\hskip -2.4mm q}+/{\hskip -2.4mm k_1}-m}
          igT_a\gamma^{\mu}\right]\right. \nonumber\\[4mm]  
        +\left.\left(\begin{array}{rl}
               k_1&\leftrightarrow~ k_2 \\[4mm]
               \mu&\leftrightarrow~ \nu
               \end{array} \right)
             \right\} 
\end{eqnarray}

\noindent Maintaining the vector current Ward indentities 
$k_{1\mu}T^{\mu\nu\lambda}=0$ and $k_{2\nu}T^{\mu\nu\lambda}=0$ we
obtain,
\begin{eqnarray}
p_\lambda T^{\mu\nu\lambda}=2mT^{\mu\nu}-
        \frac{1}{2\pi^2}\left(\frac{1}{2}\delta_{ab}g^2\kappa_{\eta'}
        \right) \varepsilon^{\mu\nu\rho\sigma}k_{1\rho}k_{2\sigma},
\end{eqnarray}
where
\begin{eqnarray}
 T^{\mu\nu}=\left(\frac{1}{2}\delta_{ab}g^2\kappa_{\eta'}\right)
          \int \frac{d^4q}{(2\pi)^4}(-1)\left\{Tr\left[
          \frac{1}{/{\hskip -2.4mm q}-m}\gamma_5 
          \frac{1}{/{\hskip -2.4mm q}+/{\hskip -2.4mm p}-m}
          \gamma^{\nu}\frac{1}{/{\hskip -2.4mm q}+/{\hskip -2.4mm k_1}-m}
          \gamma^{\mu}\right] \right.\nonumber\\[4mm]
        +\left.\left(\begin{array}{rl}
               k_1&\leftrightarrow~ k_2\\[4mm]
               \mu&\leftrightarrow~ \nu
               \end{array} \right)
             \right\} 
\end{eqnarray}

\noindent For the details of the derivation of eq.(2) and (3), 
the readers may 
refer to Ref.\cite{11}.

After a few steps of algebraic calculation eq.(3) can be finally 
converted into the following Feynman integration.
 
\begin{eqnarray}
 T^{\mu\nu}=\left(\frac{1}{2}\delta_{ab}g^2\kappa_{\eta'}\right)
              8mI(k_1^2,~k_2^2,~p^2)\varepsilon^{\mu\nu\rho\sigma}
              k_{1\rho}k_{2\sigma}, 
\end{eqnarray}
where
\begin{eqnarray}
 I(k_1^2,~k_2^2,~p^2)=\frac{-1}{16\pi^2}\int_0^1 dx\int_0^{1-x} dy
     \frac{1}{(x^2-x)k_1^2+(y^2-y)k_2^2-2xyk_1\cdot k_2+m^2}. 
\end{eqnarray}

Now some properties of eq.(2) should be pointed out. The second term
in the right hand side of eq.(2) violates 
the axial current Ward identity
$p_\lambda T^{\mu\nu\lambda}=2m T^{\mu\nu}$, it is called anomaly
term. The factor $\frac{1}{2\pi^2}\left(\frac{1}{2}
\delta_{ab}g^2\kappa_{\eta'}\right)$ is momentum independent.
Substituting eq.(4) into eq.(2) we get
\begin{eqnarray}
p_\lambda T^{\mu\nu\lambda}=
             [16m^2I(k_1^2,~k_2^2,~p^2)-\frac{1}{2\pi^2}]
          \left(\frac{1}{2}\delta_{ab}g^2\kappa_{\eta'}
        \right) \varepsilon^{\mu\nu\rho\sigma}k_{1\rho}k_{2\sigma}.
\end{eqnarray}
  
After practising the numerical calculation through the use of 
the computer we find 
for the whole range of the momentum squared $k_1^2$ and $k_2^2$, which
is favored by the decay $B\to\eta' X_s$, the first term in the square
brackets of eq.(6) is almost two orders of magnitude smaller 
than the second one.
Hence we can safely drop the first term. Now the QCD anomaly coupling
for the $\eta'-g-g$ vertex is finally given by
\begin{eqnarray}
 A^{\mu\nu}=-\frac{\alpha_s(\mu)\kappa_{\eta'}}{\pi}D\delta_{ab}
     \varepsilon^{\mu\nu\rho\sigma}k_{1\rho}k_{2\sigma}, 
\end{eqnarray}
where $D=\sqrt{3}cos\theta$ which is introduced by taking into account
the contributions of three quarks contained in $\eta'$ meson. The angle
$\theta$ is the pseudoscalar mixing angle defined by 
$\eta'=\eta_0 cos\theta+\eta_8sin\theta$ and our choice of the angle is
$\theta=-19.5^0$.

We use eq.(7) to calculate the branching ratio for 
$B\to\eta' X_s$. We use the branching ratio of $b\to sg\eta'$
to estimate the inclusive process. The strong penguin induced
$b\to s$ current is \cite{12} 
\begin{eqnarray}
\frac{G_f}{\sqrt{2}}\frac{g_s}{4\pi^2}V_t\bar{s}T_a\{\Delta F_1
   (q^2\gamma_{\mu}-q_\mu)L-F_2i\sigma_{\mu\nu}q^{\nu}m_bR\}b,
\end{eqnarray}
where $\Delta F_1\simeq -5$, $F_2\simeq 0.286$ and  
$V_t=v_{ts}^*v_{tb}$. For the convenience of expressing 
the result for the
branching ratio of $b\to sg\eta'$ we first define some variables:
$x\equiv m_x^2/m_b^2$ with $m_x$ the invariant mass
of the hadronic system in recoil from the $\eta'$ meson
in the final state,
$y\equiv k_1^2/m_b^2$ with $k_1$ the momentum of the virtual 
gluon $g^*$ which connects the penguin diagram with the $\eta'$ meson,
$x'\equiv m_{\eta'}^2/m_b^2$ and $x_s'\equiv m_s^2/m_b^2$. The 
branching ratio can be expressed as 

\begin{eqnarray}
  Br(b\to sg\eta')=
    \displaystyle\frac{1}{\Gamma_B}
    \displaystyle\frac{G_f^2|V_t|^2m_b^5}{192\pi^3}
    \displaystyle\frac{g_s^2(mb)}{16\pi^4}\displaystyle\frac{m_b^2}{4}
              \left(\frac{\alpha_s(\mu)\kappa_{\eta'}}{\pi}D\right)^2 
    ~~~~~~~~~~~~~~~~~~~~~~~~~~~~~           \nonumber\\[4mm]       
  \times\int dx\int dy \left\{\frac{1}{2}|\Delta F_1|^2[-2(x-x_x')^2y+
        (1-y-x_s')(y-x')(2x+y-x'-2x_s')] \right.\nonumber\\[4mm]
        -Re(\Delta F_1 F_2^*)(1-y-x_s')(y-x')^2/y 
    ~~~~~~~~~~~~~~~~~~~~~~~~~~~~~\nonumber\\[4mm]
 \left.+\frac{1}{2}|F_2|^2[2(x-x_s')^2 y^2-(1-y-x_s')(y-x')(2(x-x_s')y-
        (y-x')(1-x_s'))]/y^2\right\}.
\end{eqnarray}

It should be stressed that our result of the anomaly coupling for the 
$\eta'-g-g$ vertex is obtained
by assuming the divergent axial vector coupling between $\eta'$ and 
the quark line in the triangle diagram. If we only use it to explain 
one experiment, the validity of our method cannot be tested. So we 
should use the same method and the same parameter $\kappa_{\eta'}$ to
other processes. Now we will use it to calculate $\eta'
\to \gamma\gamma$ decay which is an electromagnetic decay process, so
it is completely different from $\eta'-g-g$ coupling in which strong
interaction is involved. Calculating the similar triangle diagrams as
shown in Fig.1, where the coupling $i\kappa_{\eta'}/{\hskip -2.4mm p}
\gamma_5$ is the same, only the strong coupling $igT_a\gamma^{\mu}$,
$igT_b\gamma^{\nu}$ are changed to be the QED coupling 
$iQe\gamma^{\mu}$ and $iQe\gamma^{\nu}$. We find the amplitude of 
$\eta'\to\gamma\gamma$ is
\begin{eqnarray}
 A_{\gamma\gamma}=-\frac{2\alpha\kappa_{\eta'}}{\pi}D'
        \varepsilon^{\mu\nu\rho\sigma}k_{1\rho}k_{2\sigma}
        \varepsilon^*_{\mu}\varepsilon^*_{\nu}. 
\end{eqnarray}
where $\alpha$ is the fine-structure constant, i.e., $\alpha=1/137$,
$$D'=\left[\frac{1}{\sqrt{3}}(Q_u^2+Q_d^2+Q_s^2)cos\theta +
       \frac{1}{\sqrt{6}}(Q_u^2+Q_d^2-2Q_s^2)sin\theta\right]\cdot N_c
 \simeq \frac{7}{3\sqrt{6}},$$ 
where $N_c=3$ is the color number. Finally
we get
\begin{eqnarray}
 \Gamma_{\eta'\to\gamma\gamma}=\frac{1}{64\pi}
            \left(\frac{14}{3\sqrt{6}}\frac{\alpha\kappa_{\eta'}}{\pi}
            \right)^2 m_{\eta'}^3. 
\end{eqnarray}
Comparing eq.(11) with experimental data, 
$\Gamma_{\eta'\to\gamma\gamma}^{exp.}=(4.28\pm 0.43)\times 10^{-6}$GeV
\cite{13},
we find 
\begin{equation}
\kappa_{\eta'}=7.06\pm 0.35 GeV^{-1}. 
\end{equation}

Substitute the value of $\kappa_{\eta'}$ into eq.(9), and take 
$m_b=4.8$GeV, $m_s=0.15$GeV, $\frac{1}{\Gamma_B}\frac{G_f^2|V_t|^2m_b^5}
{192\pi^3}\approx 0.2$, we can get,
 $$Br(b\to sg\eta')=(4.9\pm 0.5)\times 10^{-4}, \eqno(13)$$
here the error bar $\pm 0.5$ comes from $\pm 0.35$ in eq.(12),
and the experimental cut has been taken into account in eq.(13).
Comparing eq.(13) with the experimental data 
$Br(B\to\eta' X_s)=(6.2\pm 2.0)\times 10^{-4}$ \cite{1} 
we find that eq.(13) is fairly consistent with
the data.

Some remarks should be given. First, to get eq.(13), we have used the
running strong coupling constant $\alpha_s(\mu)$ in eq.(7)\cite{14}.
Althought it is samller than the result which is obtained 
without taking into account the running of 
$\alpha_s(\mu)$, it is well within the experiment error bar. In 
contrast, it will be too large to account for the experimental data
if we do not take into account the running of $\alpha_s(\mu)$. Second,
at $\mu=m_{\eta'}$ we find the anomaly coupling in eq.(7) is
$\frac{\alpha_s(m_{\eta'})\kappa_{\eta'}}{\pi}D\doteq 1.8GeV^{-1}$.
It is also consistent with the anomaly coupling for the $\eta'-g-g$ 
vertex which AS\cite{3} extracted from the
experimental data of $Br(J/\psi\to\eta'\gamma)$. Hence the QCD anomaly
coupling for the $\eta'-g-g$ vertex derived by calculating the 
triangle diagram can
also give the correct result of $Br(J/\psi\to\eta'\gamma)$. Third, in
general, it is believed that the QCD anomaly coupling of $\eta'-g-g$
is nonperturbative and hence it is unpredictable. But now our 
calculation shows that the nonperturbative part can be successfully 
separated, which is obsorbed into the coupling parameter $k_{\eta'}$. 
With the parameter 
$k_{\eta'}$ the strong coupling of $\eta'-g-g$ can be predicted 
from the knowledge of
$\eta'\to \gamma\gamma$, which is an electromagnetic decay process.
It is very interesting to note that the QCD anomaly and QED anomaly 
can be treated in an uniform way.

In summary, by calculating AVV (Axial vector-Vector-Vector current)
triangle diagram, we find that QCD anomaly coupling for the $\eta'-g-g$
vertex and QED anomaly coupling $\eta'-\gamma-\gamma$ can be treated 
with one parameter $\kappa_{\eta'}$. The calculation of these three
decay process $B\to\eta' X_s$, $J/\psi\to\eta'\gamma$, and 
$\eta'\to \gamma\gamma$ can be consistent with experiment at the same
time. SM is sufficient to account for $B\to\eta' X_s$ data within the 
present experimental error bars.
     
The outhors would like to thank Dr. C. D. L\"{u} and Dr. T. Morozumi 
for useful discussions. One of us (MZY) thanks Japan Society for the 
Promotion of Science (JSPS) for financial support.


\begin{thebibliography}{99}
\bibitem{14} The running of the effective QCD coupling constant is 
             $$\alpha_s^{(f)}(\mu)=\frac{4\pi}{\beta_0
              ln(\mu^2/\Lambda_f^2)}\left[1-\frac{\beta_1}{\beta_0^2}
              \frac{lnln(\mu^2/\Lambda_f^2)}{ln(\mu^2/\Lambda_f^2)}
              \right]$$
              with $\beta_0=11-\frac{2}{3}f$ and 
              $\beta_1=102-\frac{38}{3}f$, where $f$ is the number of
              the quarks which are active in the scale of $\mu$. 
              $\Lambda_4=300MeV$ is used in this paper, and to keep the 
              running of $\alpha_s^{(f)}(\mu)$ continuous at 
             $\mu=m_c=1.5GeV$, $\Lambda_3=353MeV$ is taken here.
\end{thebibliography}
\end{document}